\documentclass[12pt]{article}
\usepackage[dvips]{epsfig}
\makeatletter
\def\lsim{\mathrel{\mathpalette\@versim<}}
\def\gsim{\mathrel{\mathpalette\@versim>}}
\def\@versim#1#2{\vcenter{\offinterlineskip
    \ialign{$\m@th#1\hfil##\hfil$\crcr#2\crcr\sim\crcr } }}
\makeatother

\begin {document}
\thispagestyle{empty}
\begin{center}
{\Large \bf Adsorption and collapse transitions of a linear
polymer chain interacting with a surface adsorbed polymer chain}
\end{center}

\begin{center}
{\large \bf Sanjay Kumar}
{\large \bf and Yashwant Singh} \\ 
{\em \bf  Department of Physics, Banaras Hindu University} \\
{\em \bf  Varanasi 221 005 India.}
\end{center}
\date{today} 

\begin{abstract}
We study the problem of adsorption and collapse transition of a
linear polymer chain situated in a fractal container represented
by a 4-simplex lattice and interacting with a surface adsorbed
linear polymer chain.  The adsorbed chain monomers act as
pinning sites for the polymer chain.  This problem has been
solved exactly using real space renormalization group
transformation. The resulting phase diagram and critical
exponents are given. 
\end{abstract}

\vspace {1in}

\noindent {\bf PACS :} 64.60.Ak, 64.60.Fr, 64.60.Kw \\
\noindent {\bf Keyword :} Interpenetration, Contact Exponent, Polymer, 
Fractal.

\newpage

\section{Introduction}
The problem of surface effects on conformation statistics of
long flexible polymer chains has been widely studied both
because of its merit as an interesting problem in statistical
mechanics and because of its important role in many physical
processes like colloidal stabilization, adhesion, or
lubrication, etc \cite{1,2}. The statistical mechanics approach
to this problem has successfully been applied, particularly in
case of good solvent that contains only one linear polymer chain
interacting with an impenetrable wall \nocite{3,4,5,6,7} [3-7]. 
The essential physics is derived from a model of self-avoiding
walk (SAW) on a semi-infinite lattice, with an energy 
contribution $\epsilon_a$ for each step of the walk along the
lattice boundary. This leads to an increased probability
characterized by the Boltzmann factor $\omega = \exp
(-\epsilon_a / k_{\beta}T)$ of making a step along the
attractive wall, since $\epsilon_a < 0$, $\omega > 1$ for any
finite temperature $T$. At low temperatures due to the attraction
between the polymer chain and the surface, the chain  gets adsorbed 
on the surface while at
high temperatures all polymer configurations have almost same
weight and a non-adsorbed behaviour prevails. The transition
between these two regions is marked by a critical adsorption
temperature $T_a$, with a desorbed phase for $T > T_a$ and an
adsorbed phase for $T < T_a$. The asymptotic behaviour of the
average number $M$ of steps of the walk along the boundary can
be summarized in the following way \cite{7}
\begin{eqnarray}
M \sim \left \{ \begin{array}{ll}
(T_a - T)^{\frac{1}{\phi} -1}  &  T < T_a  \\
N^{\phi} & T = T_a \\
(T - T_a)^{-1} &  T > T_a 
\end{array}
\right.
\end{eqnarray}
where $N$ denotes the average number of monomers and $\phi$ is
the crossover exponent.

When the polymer chain is in a poor solvent, it exhibits a phase
diagram characterized by many different universality domains of
critical behaviour. This is due to the competition between solvent -
induced monomer-monomer attraction and the surface-monomer
interaction. In this case the essential physics is derived using
a model of self-attracting self-avoiding walk (SASAW) on a
semi-infinite lattice \cite{8,9}.

Theoretical methods which have been used to study polymer adsorption
include renormalization group \cite{10,11}, transfer matrix
\cite{12}, Monte Carlo \cite{13,14}, exact enumeration \cite{15}
and series expansion techniques \cite{16}. In case of two
dimensions many exact results have been found through conformal
field theory [17-20] \nocite{17,18,19,20} and using conformal
invariance prediction in conjunction with the Bethe ansatz
solution of associated lattice models \cite{21,22}. Many exact
results have also been found for the case of fractal lattices
using real space renormalization group (RSRG) [8, 23-25]
\nocite{8,23,24,25}. In this article, we consider the adsorption 
and collapsed transition of a long flexible polymer chain interacting 
with another long flexible polymer chain adsorbed on a surface. 
The monomers of the adsorbed chain act as pinning (or interaction) 
sites for the chain of our interest. The situation is shown in 
Figure 1. A polymer chain shown by zigzag line lies on the surface.
Its monomers shown by black circles act as a pinning sites for 
the other polymer chain which is shown by wiggle line. The monomers 
of this floating chain shown by open circles get attracted to the 
surface because of the pinning sites. When a monomer of the floating
chain gets adsorbed on the pinning sites (shown by shade circle) 
there is a gain of energy $\epsilon_a$. The model which we describe 
below also has two other interactions; (i) attraction between nearest
neighbours formed by monomers of floating chain which are separated 
by relatively large distance along the chain and (ii) the interaction 
between nearest neighbours formed by a pinning site and a monomer of 
the floating chain occupying the neighbouring sites. These interactions 
are shown by $\epsilon_u$ and $\epsilon_t$, respectively, in the
Figure 1.  This situation is similar to the problem 
of adsorption of a long flexible polymer chain onto a (cell) 
membrane along which protein stick out in a spatially uncorrelated 
manner.

The paper is organized as follows. In section 2, we describe a
model of two interacting chains in which one is adsorbed on a 
surface of a fractal container. The motivation of
choosing a fractal space is that the model can be solved exactly
apart from having its own practical applications. The results
found by solving the model exactly are given in section 3. The
paper ends with conclusions listed in section 4. 

\section{Model and its solution}

We consider a fractal space mapped by truncated 4-simplex
lattice. The basic geometrical unit of construction of this
lattice is a tetrahedron with 4-corner vertices and bonds
between every pair of vertices. Each vertex connected through a
direct bond is termed as nearest neighbour. The fractal and
spectral dimensions of the 4-simplex lattice are 2 and 1.5474,
respectively. The tetrahedron of first and $(r+1)$th order are
shown in Figure 2. The shaded region represents the surface. 
The surface is a truncated 3-simplex lattice with fractal and spectral
dimensions, 1.5849 and 1.3652 respectively. The bulk critical
behaviour including $\theta$-point and the phase diagram of
surface interacting polymer chain using SASAW model has been
studied recently \cite{8}. 

In the model proposed here we represent the pinning sites by
the monomers of an adsorbed polymer chain on a 3-simplex
lattice. The configuration of the pinning sites can, therefore,
be found by the statistics of single polymer chain on the
3-simplex lattice. For convenience we represent this
configuration (of adsorbed polymer chain) by $P_2$. The floating 
polymer chain in a fractal container whose adsorption we want to 
study is represented by $P_1$. Therefore the problem is projected on 
to a model of two interacting crossed walks [26-28] \nocite{26,27,28} 
in which one walk is confined to the surface and act as pinning sites
for the other chain. 

We assign the weight $x_1 (x_2)$ to each step of the walk in the
bulk (on the surface) and the weight $\sqrt{x_1 x_2} \omega$ to
each step taken on the pinning sites on the surface. In other
words, when a monomer of chain $P_1$ visits a site on the surface
occupied by chain $P_2$ a weight $\sqrt{x_1 x_2} \omega$ is
assigned. If a monomer visits a site on the surface not occupied by
the chain $P_2$ ({\it i.e} a pure site) the weight assigned to it is
$x_1$. The monomers of chain $P_1$ may attract each other. We
denote the Boltzmann factor associated with this interaction by
$u$, where $u = \exp (-\epsilon_u/k_{\beta}T)$ ($\epsilon_u < 0$
being the attractive energy associated with a pair of
near-neighbour bonds on the lattice). In order to promote
competition between the adsorbed and desorbed phases of chain
$P_1$ it is desirable to introduce a parameter $t = \exp
(-\epsilon_t/k_{\beta}T)$ in such a way that $\sqrt{x_1 x_2} t$
is the weight of those steps that are performed on the lattice
points which are the nearest neighbour but in a adjacent sites 
to the pinning sites. Here $\epsilon_t$ is the interaction energy 
between a pair formed by a monomer of chain $P_1$ in the
adjacent site and a pinning site on the surface. 

The global generating function of this model can be written in
the form 
\begin{eqnarray}
\Omega(x_{1},x_{2},{\omega},u,t) & = &\sum_{all\;walks}x_{1}^{N_{1}}
x_{2}^{N_{2}}{\omega}^{N_{s}}t^{N_{c}}u^{N_{m}} \\ \nonumber
& =  & \sum_{N_{1} N_{2} N_{s} N_{c} N_{m}} C(N_{1}N_{2}N_{s}N_{c}N_{m})
x_{1}^{N_{1}}x_{2}^{N_{2}}{\omega}^{N_{s}}t^{N_{c}}u^{N_{m}}
\end{eqnarray}
where $C(N_{1},	N_{2},N_{s},N_{c},N_{m})$ represents the total 
number of configurations of all walks. Here $N_{1}$ is the total
number of monomers in chain $P_{1}$, $N_{2}$ represents the total 
number of monomers in adsorbed polymer chain ($P_{2}$), 
$N_{s}$ denotes the number of monomers of polymer chain $P_{1}$ 
adsorbed on the pinning sites. $N_{c}$ and $N_{m}$ are number of 
monomers lying adjacent to the surface and forming the nearest 
neighbour with the pinning sites, and number of nearest neighbours
in chain $P_1$, respectively. 

The generating function for the 4-simplex lattice can be
expressed in terms of finite number of restricted partition
functions \cite{29}. We show in Figure 3 all the possible restricted
partition functions which appear in this problem. This can be
seen from Figure 4 in which we draw one of the possible
configurations of walks on the $r$-th order of the lattice. The
surface adsorbed chain $P_2$ is shown by zigzag line while the chain
$P_1$ by wiggle line. These partition functions are defined
recursively as weighted sum over all possible configurations for
a given stage of the iterative construction of the 4-simplex
lattice. The variables in these equations are just the partial
generating functions corresponding to different polymer
configurations for a given size of the fractal. Linearizing the
recursion equations near the fixed points, the one reached by the system
depending on the initial conditions, we can find the eigenvalues
of the transformation matrix which give the characteristic
exponents of the system.

The recursion relations for the restricted partition functions
can be written as (see Figure 5)
\begin{eqnarray}
A_{r+1} & = & A^{2}+2A^{3}+2A^{4}+4A^{3}B+6A^{2}B^{2} \\
B_{r+1} & = & A^{4}+4A^{3}B+22B^{4} \\
H_{r+1} & = & H^{2}+H^{3} \\
C_{r+1} & = & C^{2}+C^{3}+AD^{2}(H+2C+2F+2G+2+2A)+\\ \nonumber 
&  &  4A^{2}G(C+F)+2AG^{2}(A+C)+4ACFG+HD^{2} \\
D_{r+1} & = & ABD(4G+2F+2E)+A^{2}D(G+F+E+C+H)+\\  \nonumber 
&  & AD(C+H+C^{2}+H^{2})+BDE(2F+4G)+\\ \nonumber 
&  &ACD(F+G+H)+AHDE+D^{3}(2B+A)  \\
E_{r+1} & = & AD^{2}(4B+2G+2E+2H+2A)+BD^{2}(2F+4G)+\\ \nonumber 
&  & AH^{2}(A+E)+6B^{2}E^{2}+2BE^{3}+2A^{2}HE \\
F_{r+1} & = & AD^{2}(2B+E+2G)+B^{2}(8G^{2}+6F^{2}+\\  \nonumber 
&  & B(8G^{3}+2F^{3})+8BFG(B+G)+ACF(2A+C)+\\  \nonumber 
&  &BED^{2}+4BGF^{2}+A^{2}C^{2} \\
G_{r+1} & = & AD^{2}(2B+A+G+F+C)+AG(2AC+C^{2})+\\  \nonumber  
&  &12B^{2}FG + 6BGF^{2}+10BFG^{2}+10B^{2}G^{2}+\\  \nonumber 
&  &6BG^{3}+BD^{2}E
\end{eqnarray}

A notational simplification in which the index $r$ is dropped from
the right hand side of the recursion relations is adopted here. 
It may be emphasized here that the recursion relations written above 
are exact for the model defined above. 

Eq.(2.5) which represents the recursion relation for the
adsorbed chain $P_2$ is independent of configuration of
chain $P_1$. The effect of $P_2$ on chain $P_1$ is taken through
$C$, $D$, $E$, $F$ and $G$. Since all interactions involved in
the problem are restricted to bonds within a first order unit of
the fractal lattice, $\omega$, $t$, and $u$ do not appear
explicitly in the recursion relations. They appear only in
initial values given below. 
\begin{eqnarray}
A_{1} & = & x_{1}^{2}+2x_{1}^{3}u+2x_{1}^{4}u^{3} \\
B_{1} & = & x_{1}^{4}u^{4} \\
C_{1} & = & x_{1}^{2}x_{2}^{3}{\omega}^{2}t^{2} +
x_{1}^{3}x_{2}^{3}({\omega}^{6}u+t^{2}{\omega}^{2}u) +
2x_{1}^{4}x_{2}^{3}{\omega}^{4}t^{2}u^{3} +
x_{1}^{2}x_{2}^{2}{\omega}^{4}+ \\ \nonumber  
& & 2x_{1}^{3}x_{2}^{2}t^{2}{\omega}^{2}u +
2x_{1}^{4}x_{2}^{2}t^{2}{\omega}^{2}u^{3} \\
D_{1} & = & x_{1}^2x_{2}^{3}t{\omega}+x_{1}^{3}
x_{2}^{3}(t{\omega}u + t{\omega}^{3}u) + x_{1}^{4}x_{2}^{3}
(u^{3}{\omega}^{3}t^{3}+u^{3}{\omega}^{5}t) + \\ \nonumber 
& &x_{1}^{2}x_{2}^{2}t{\omega}+ x_{1}^{3}x_{2}^{2}({\omega}^{3} 
tu+{\omega}tu)+x_{1}^{4}x_{2}^{2}({\omega}^{3}tu^{3}+{\omega}tu^{3}) \\
E_{1} & = & x_{1}^{2}x_{2}^{3}+2x_{1}^{3}x_{2}^{3}{\omega}^{2}
t^{2}u+2x_{1}^{4}x_{2}^{3}t^{2}{\omega}^{2}u^{3} +
x_{1}^{2}x_{2}^{2}+2x_{1}^{3}x_{2}^{2}u+\\ \nonumber 
&  &2x_{1}^{4}x_{2}^{2}u^{3}t^{2}{\omega}^{2} \\
F_{1} & = & x_{1}^{4}x_{2}^{2}{\omega}^{4}u^{4} +
x_{1}^4x_{2}^{3}{\omega}^{2}t^{2} u^{4} \\
G_{1} & = & x_{1}^{4}x_{2}^{2}t^{2}{\omega}^{2}u^{4} +
x_{1}^{4}x_{2}^{3}{\omega}^{4}t^{2}u^{4} 
\end{eqnarray}

Here index 1 and 2 on the right hand side of Eq.(2.11)-(2.17)
correspond to chain $P_{1}$ and pinning sites (chain $P_{2}$)  
respectively. The fixed points corresponding to different
configurations of polymers in the asymptotic limit are found by
solving Eq.(2.3) and (2.4) for polymer chain $P_{1}$ and 
Eq.(2.5) for chain $P_{2}$ respectively. A complete phase 
diagram obtained from Eqs.(2.3) - (2.4) and from Eq.(2.5) are
given elsewhere \cite{8}. 

The state of polymer chain $P_{1}$ depends on the quality of the
solvent and on the temperature and can therefore be in any of
three states; swollen, compact globule and at $\theta$-point
described in the asymptotic limit by the fixed points $
(A^{*},B^{*})$ = (0.4294, 0.0498), (0.0,$22^{-1/3})$ and
(1/3,1/3) respectively. The fixed point corresponding to the
swollen state is reached for all values of $u < u_{\theta}$ at 
$x_1 > x_{\theta}$. The end to end distance for a chain 
of $N_{1}$ monomers of $P_{1}$ in this state varies as
$N_{1}^{\nu_{1}}$ with $\nu_{1}$ =0.7294 \cite{8,29} and
connectivity constant $\mu = 1/x$ is found to be 1.5474.
The fixed point corresponding to the compact globule state is 
reached for all values of $ u > u_{\theta}$ at 
$x_{1}(u) < x_{\theta}$. At $u_{\theta}$ = 3.31607.. and 
$x_{\theta}$ = 0.22913.. the system is found to be at its
tricritical point or $\theta$-point. The fixed point $H^{*}$ =
0.61803.. is found by solving Eq.(2.5). It corresponds to a
pattern with fractal dimension equal to 1.266.

In a system of polymer chain interacting with a surface adsorbed 
chain, we have three different combinations of the individual state. 
Using the fixed points of $(A^{*},B^{*})$ corresponding to
three different states of polymer chain $P_{1}$ and $H^{*} = 0.61803$ 
(for surface adsorbed chain $P_2$), we solve the coupled
non-linear equations [Eqs. (2.6) - (2.10)].

\section{Results}

For all values of $\omega < \omega_c (u,t)$, the polymer chain $P_1$ 
lies in the bulk. The critical behaviour of polymer chain $P_1$
does not get affected by the presence of the surface or pinning 
sites. In the phase diagram shown in Figure 6, $\omega$
is plotted as a function of $u$ for three values of $t = 0$, 0.5
and 1.0. The $\theta$-line which separates the bulk swollen and
collapsed phases and terminates at the surface adsorption line 
$\omega = \omega_c (u,t)$ is found at $u = u_c = 3.316074$. The
$\theta$-line in Figure 6 is shown by dashed line. $\omega =
\omega_c (u,t)$ separates the bulk from the adsorbed phase.
Below this line the desorbed phase does not get affected by the
presence of the surface attraction, except for the
$\theta$-point turning into a $\theta$-line and the critical
lines corresponding to swollen and collapsed phases into
respective regions. 

When $\omega > \omega_c (u,t)$ the polymer gets adsorbed.
Depending on the value of $t$ the chain may lie on the
pinning sites or avoids them. For $t \sim 1$ the polymer
chain may get adsorbed on the pinning sites acquiring the
configuration as that of the pinning sites. 

When $\omega = \omega_c (u,t)$ the chain $P_1$ is on the adsorption
special line. The different regions of this line characterize
different multicritical behavior as a function of $u$ and $t$.
Note that the parameter $u$ and $t$ measure, respectively, the
strengths of nearest neighbour interaction and the repulsive
strength of a monomer on the adjacent layer  and forming a 
nearest neighbour with a pinning site on the surface. 

\par (I) The fixed point $A^{*}$, $B^{*}$,$C^{*}$, $D^{*}$,
$E^{*}$, $F^{*}$, $ G^{*}$ ) = (0.4294, 0.04998, 0.2654, 0.2654,
0.2654, 0.0391, 0.0391) is reached for all values of $u < u_{c}$
and $\omega = \omega_{c}$ at $t > 0$. Linearization around this
fixed point gives one eigenvalue (other than $\lambda_{b} =
2.7965$) $\lambda_{\phi} = 1.7914$ greater than one. From the
eigenvalues $\lambda_b$ and $\lambda_{\phi}$ one gets the
crossover exponent 
\begin{displaymath}
\phi = \frac{\ln \lambda_{\phi}}{\ln \lambda_b} = 0.5669 \sim
0.57 
\end{displaymath}

\noindent This fixed point corresponds to special line $\omega =
\omega_{c} (u,t)$ for $u < u_c$ and $t > 0$. The polymer chain
at the adsorption line (tricritical line) fluctuates among
configurations corresponding to $A$ (bulk), $C$, $D$, $E$
(surface) [see Figure 3]. 

For $t=0$ the fixed point ($A^{*}$, $B^{*}$, $C^{*}$,
$D^{*}$, $E^{*}$, $F^{*}$, $G^{*}$ ) =(0.4294, 0.04998, 0.0, 
0.0, 0.1164, 0.0, 0.0) is reached for all values of $u < u_{c}$.
Linearization about this fixed point does not yield any
eigenvalue (except $\lambda_b$) greater than one. Therefore
there is no crossover from bulk to surface.

\par (II) When $u > u_{\theta}$ (= 3.316074) and $x <
x_{\theta}$ (= 0.229157...) the polymer chain is in the
collapsed state in the bulk. At $t = 0$, we choose $u = 3.3333..$ and
$x = 0.2282..$ and solve Eqs.(2.3) - (2.9). This leads to a fixed
point ($A^{*}, B^{*}, C^{*}, D^{*}, E^{*}, F^{*}, G^{*}$) =
(0.0, 0.3568, 0.0, 0.0, 0.7652, 0.0, 0.0) at $\omega = \omega_c
= 1.5751....$. This corresponds to configuration in which
compact globule formed in the bulk container gets stuck to the
surface in such a way that monomers avoid touching the sites of the 
surface occupied by the chain $P_2$. Linearization around this fixed point
gives in addition to $\lambda_b$ an eigenvalue greater than one
{\it i.e} $\lambda_{\phi} = 2.4233$. With this eigenvalue the
crossover exponent is found to be 
\begin{displaymath}
\phi = \frac{\ln \lambda_{\phi}}{\ln \lambda_b} \simeq 0.64.
\end{displaymath}

\par (III) When $t > 0$ and $\omega = \omega_c$ (for $u >
u_{\theta}$ and $x < x_{\theta}$), we get a fixed point ($A^{*},
B^{*}, C^{*}, D^{*}, E^{*}, F^{*}, G^{*}$) = (0.0, 0.3568, 0.0,
0.0, 0.0, 0.2206,0.2206). Linearization around this fixed point
gives $\lambda_{\phi} = 2.3073$. The crossover exponent in this
case is $\phi \simeq 0.60$. In this case the surface
configurations attained by the polymer chain are those which
correspond to the configurations $F$ and $G$ shown in Figure 3. 
It means that the globule gets attached to the site occupied by 
the chain $P_2$ on the surface.

\par (IV) When $u = u_{\theta} = 3.316074.....$ and $x =
x_{\theta} = 0.229137...$ the polymer chain (in bulk) is at
$\theta$-point. For this we have three distinct fixed points
depending on the values of $t$. We now discuss these three fixed
points.  
\par (A) For $t= 0$ and $\omega = \omega_c = 1.5772$ the fixed
point achieved by the system is ($A^{*}$, $B^{*}$, $C^{*}$,
$D^{*}$, $E^{*}$, $F^{*}$, $G^{*}$) = (1/3, 1/3, 0.0, 0.0,
0.613, 0.8228.., 0.0). Linearized equations around this fixed
point give three eigenvalues greater than one. These values are 
\begin{displaymath}
\lambda_1 = 2.4511, 
\end{displaymath}
and 
\begin{displaymath}
\lambda_{b1} = 3.7037, \; \; \; \; \lambda_{b2} = 2.2222
\end{displaymath}
Note that the last two eigenvalues are the same as those found
for the bulk $\theta$-point. The crossover exponent is
\begin{displaymath}
\phi= \frac{\ln \lambda_1}{\ln \lambda_{b1}} \sim 0.65.
\end{displaymath}

This is a tetracritical point. The adsorbed phase corresponding
to configurations which forms a layer on the surface occupied by
the pinning sites.

\par (B) For $0 < t < 1$ we have fixed point ($A^{*}$, $B^{*}$,
$C^{*}$, $D^{*}$, $E^{*}$, $F^{*}$, $G^{*}$) = (1/3, 1/3,
0.0510.., 0.0, 0.613, 0.2364.., 0.2364). This point has been
found to have three eigenvalues greater than one. These values
are 
\begin{displaymath}
\lambda_1 = 2.3311,
\end{displaymath}
and the other two are $\lambda_{b1}$ and $\lambda_{b2}$ given
above. The crossover exponent, $\phi$ found in this case is
equal to 0.61. This tetracritical point differs from the
previous one in the sense that the adsorbed phase has
configuration which is combination of both $F$ and $G$ as shown 
in Figure 3 whereas at $t = 0$ the adsorbed phase has the
configuration corresponding to $E$. 

\par (C) At $t = 1$ and $\omega = 1$ we find the symmetrical
fixed point ($A^{*}$, $B^{*}$, $C^{*}$, $D^{*}$, $E^{*}$,
$F^{*}$, $G^{*}$) = (1/3, 1/3, 0.2061, 0.2061, 0.2061, 0.2061,
0.2061). This point has four eigenvalues greater than
one. Apart from the two known eigenvalues ($\lambda_{b1}$,
and $\lambda_{b2}$) the two additional eigenvalues are 
\begin{displaymath}
\lambda_{1} = 2.1269 \; \;{\rm and} \; \; \lambda_{2} =
1.1947 
\end{displaymath}

This is a pentacritical point. Note that for $t > 1$, corresponds to 
$\omega < 1$. Therefore the tetracritical line as a function of
$t$ is symmetrical about the point $t = 1$. 

\section{Conclusions}

In this paper we studied the critical behaviour of polymer chain
interacting with a surface adsorbed chain. It is shown that this model 
differs from the usual polymer adsorption and also from the problem 
of two interacting chains studied in past [15,23-28]
\nocite{15,23,24,25,26,27,28}.  We find a very 
rich $\omega-u$ phase diagram plotted in Figure 6. The adsorbed 
polymer chain representing the pinning sites always remain in 
swollen state with radius of gyration exponent equal to that of
truncated 3-simplex lattice. We therefore have pinning sites
forming pattern with fractal dimension 1.266. The bulk desorbed
phase has two regions: the region of swollen state separated
from the collapsed globule state (by a tricritical
$\theta$-line). The $\theta$-line is at $u = u_{\theta}$ =
3.316074.. and runs parallel to the $\omega$ axis {\it i.e.}
remains unaltered due to the surface interaction. The point
where it meets the line $\omega^{*}(t,u)$ is a multicritical
point. These multi-critical points are characterized by three
different fixed points depending on the value of $t$. When the
value of $t=1$ the value of $\omega$ is found to be 1. The
$\omega$ line runs parallel to the $u$-axis and meets at the
$\theta$-point. This point has four eigenvalues greater than one
and corresponds to the {\it pentacritical point}. This is a point
at which two tetracritical lines corresponding to $0 < t \leq 1$
and $t > 1$ meet. When $t >1$, $\omega$ has to be less than $1$.

When $t=0$, the value of surface interaction increases with $u$
and meet at $\theta$-line. The Figure 6 gives the impression of the
existence of reentrant adsorbed phase as $u$ is increased. One
should, however, remember that these figures are merely a
projection on the $\omega-u$ plane of three dimensional figures
in which the third dimension is given by $x$. 

When the value of $t$ lies in between 0 and 1 the slope
increases, but we do not find any ``frustrated phase" \cite{8} 
as observed in usual situation of surface adsorption.
The behaviour of special adsorption line described above
can be understood from contributions of different coexisting
polymer configurations (see Figure 3) to the bulk and surface free
energies. When both adsorbed phase (as by definition) and bulk
phase are in swollen state, the adsorption line has same nature
in $\omega-u$ plane for all values of $t$, although the slope
of line decreases as $t$ increases and at $t=1$ the slope
becomes zero and line runs parallel to $u$ axis. This is due to
the very fact that at $t=1$ and $\omega = 1$ the surface behaves
as a part of bulk and distribution of monomers are isotropic.

\vspace {.2in}

\noindent {\large \bf Acknowledgement}

\vspace {.2in}

We thank the Department of Science and Technology of Govt. of
India for financial assistance. We also thank D. Dhar for many
helpful discussions.

\newpage 

\begin{center}
{\bf \large FIGURE CAPTION}
\end{center}

\noindent {\bf Figure 1} Schematic   representation of all 
possible interactions  appearing in Generating Function
defined by Equation 2.2. 

\vspace {.2in}

\noindent {\bf Figure 2} Graphical representation of a truncated
4-simplex lattice of first order and ($r+1$)th order. The shaded regions
represent the surface.

\vspace {.2in}

\noindent {\bf Figure 3} Diagrammatic representation of all the
restricted partition functions which appear in the generating
function defined by Eq.(2.2). Polymer chain $P_1$ in the bulk is
represented by wiggle line while adsorbed polymer chain $P_2$ is 
shown by zigzag line.

\vspace {.2in}

\noindent {\bf Figure 4} Diagrammatic representation of one of
possible configurations of a polymer interacting with surface
confined polymer chain on third order of $4$-simplex lattice.
All the possible partition functions contributing to the
generating function are shown. 

\vspace {.2in}

\noindent {\bf Figure 5} Diagrams representing the recursion relation for
the restricted partition functions given by Eqs. 2.3-2.9. Some of the
possible configurations contributing to the recursion relations are shown.

\vspace {.2in}

\noindent {\bf Figure 6} Special adsorption lines are shown in  $\omega$ 
 -$u$ plane for $t$ = 0, 1 and 0.5.

\newpage

\begin{figure}[H]
\begin{center}
\epsfig{file=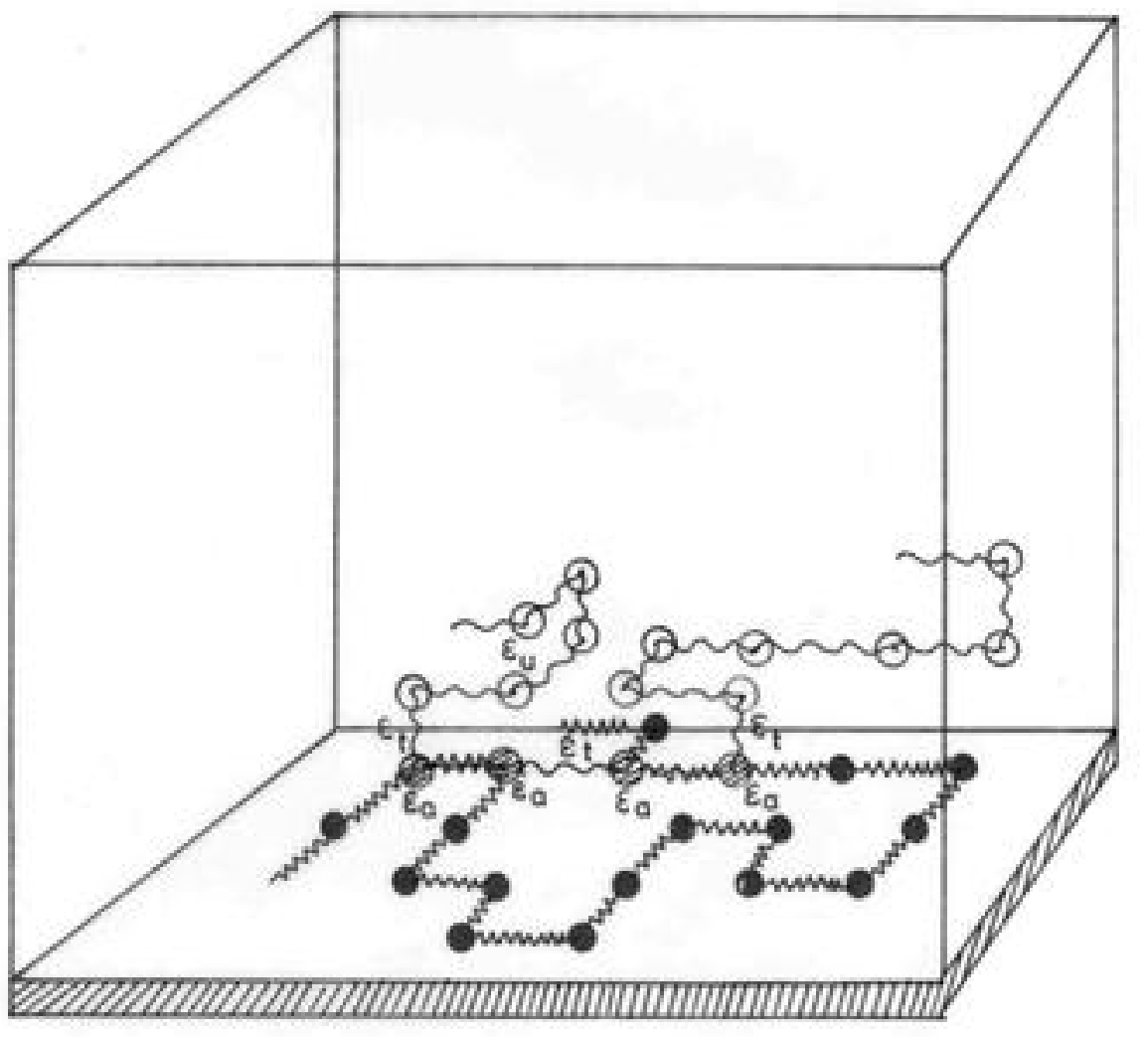,width=12cm}
\caption{ }
\end{center}
\end{figure}
\newpage
\begin{figure}[H]
\begin{center}
\epsfig{file=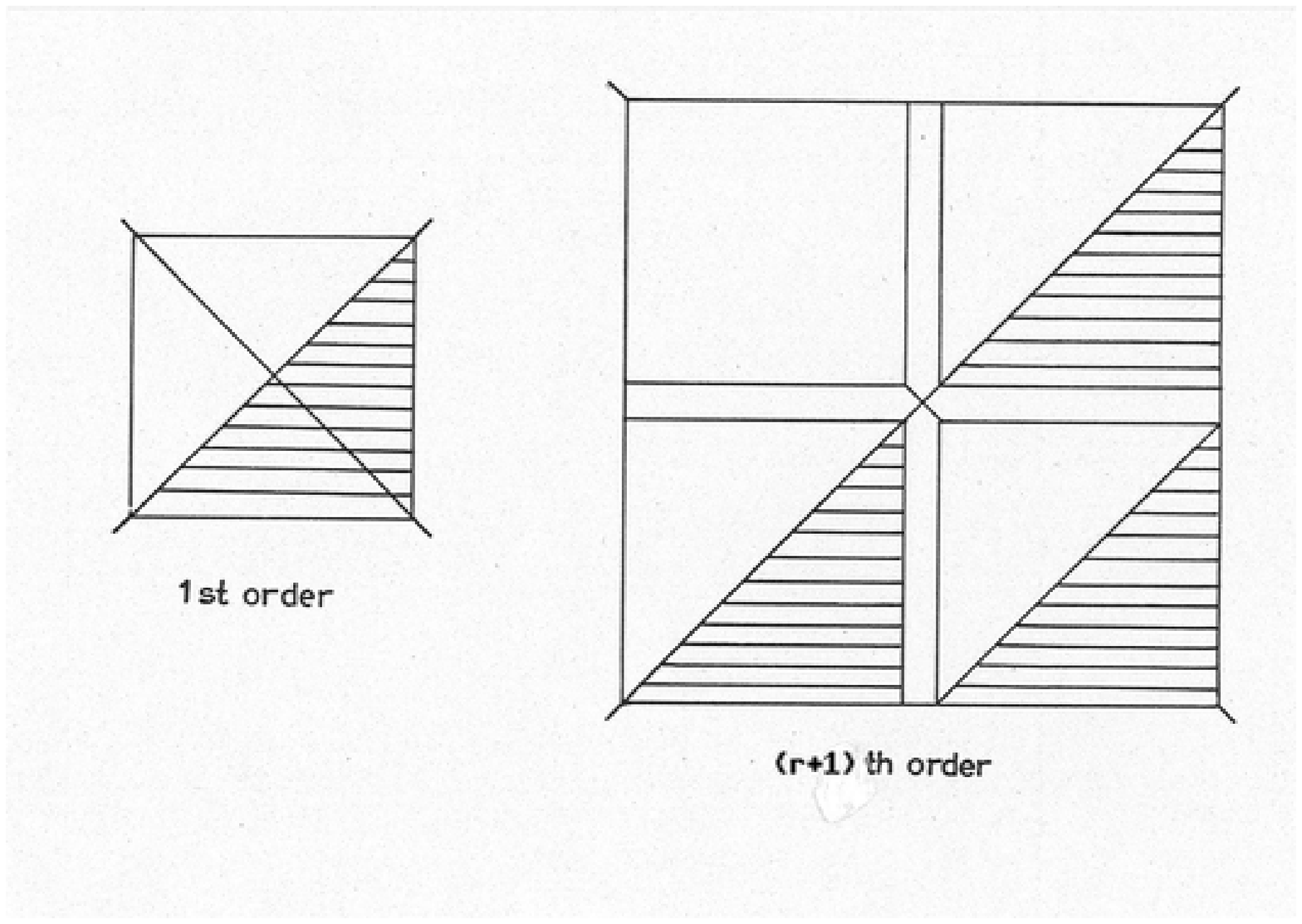,width=12cm}
\caption{ }
\end{center}
\end{figure}
\newpage
\begin{figure}[H]
\begin{center}
\epsfig{file=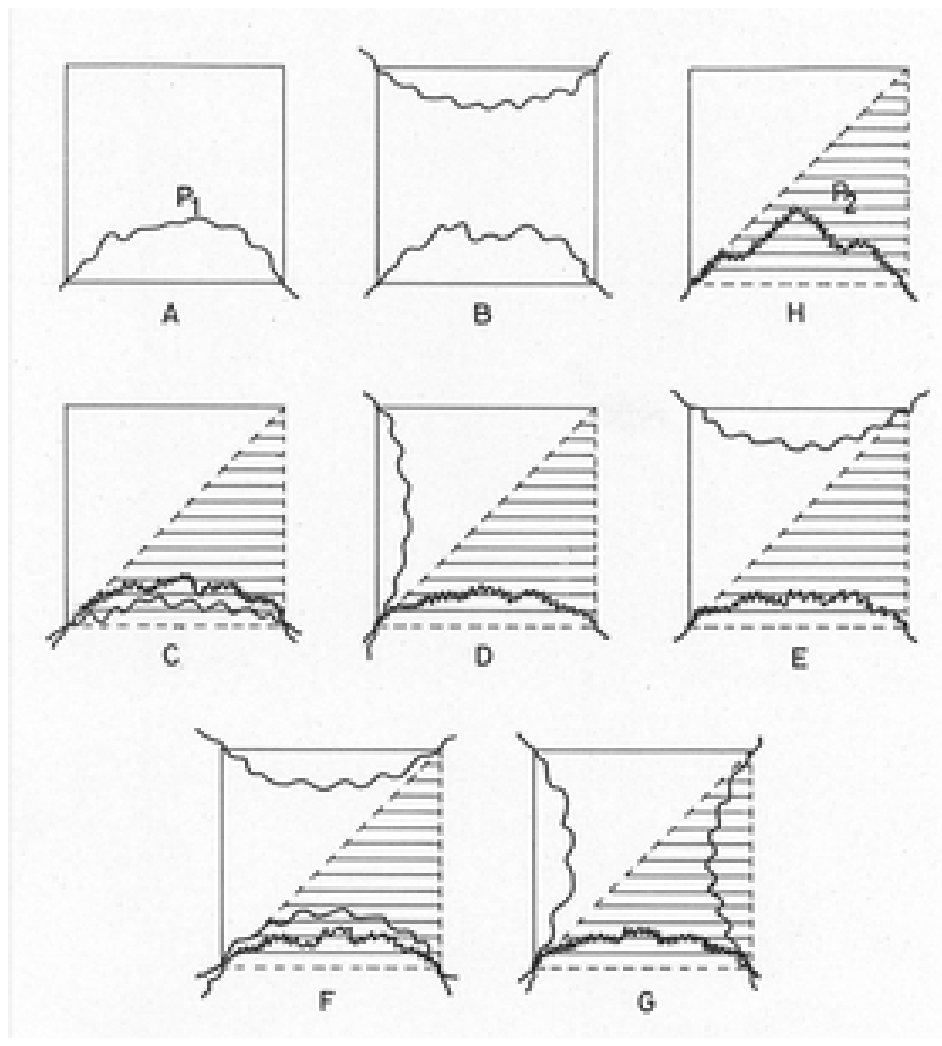,width=12cm}
\caption{ }
\end{center}
\end{figure}
\newpage
\begin{figure}[H]
\begin{center}
\epsfig{file=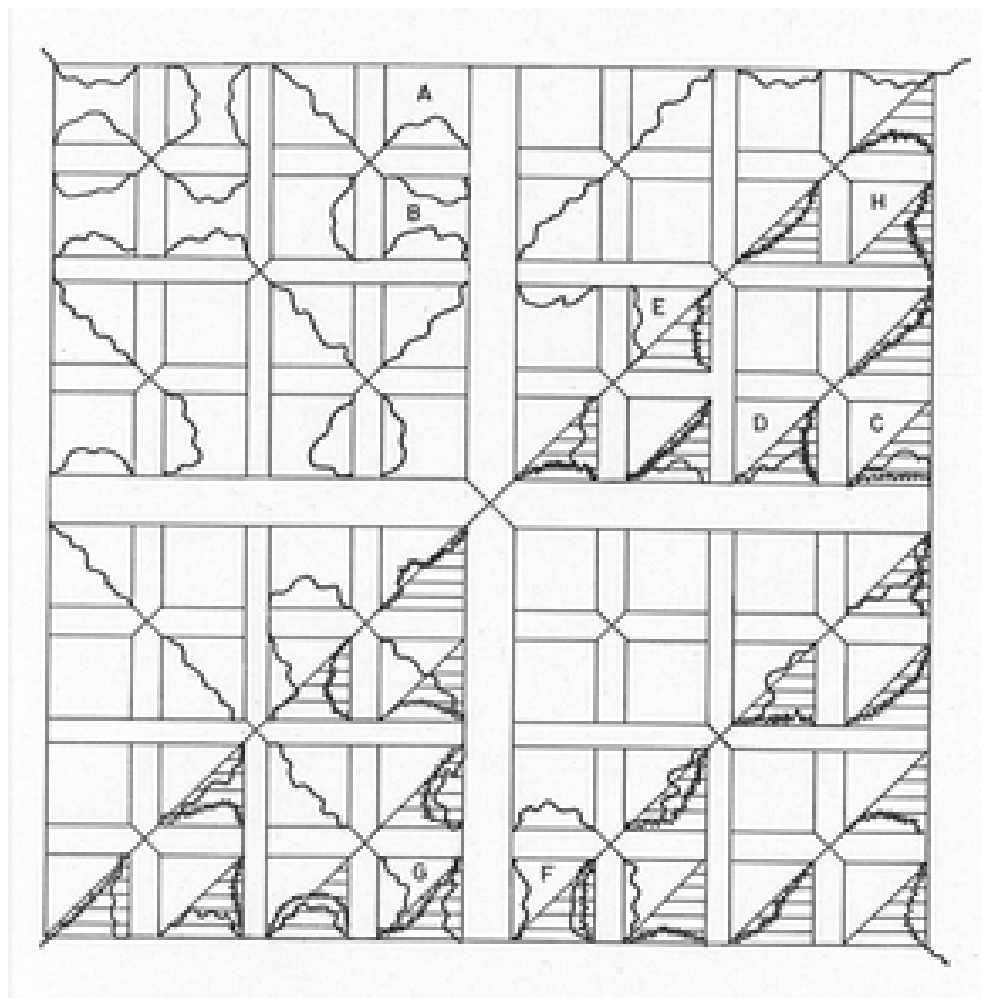,width=12cm}
\caption{ }
\end{center}
\end{figure}
\newpage
\begin{figure}[H]
\begin{center}
\epsfig{file=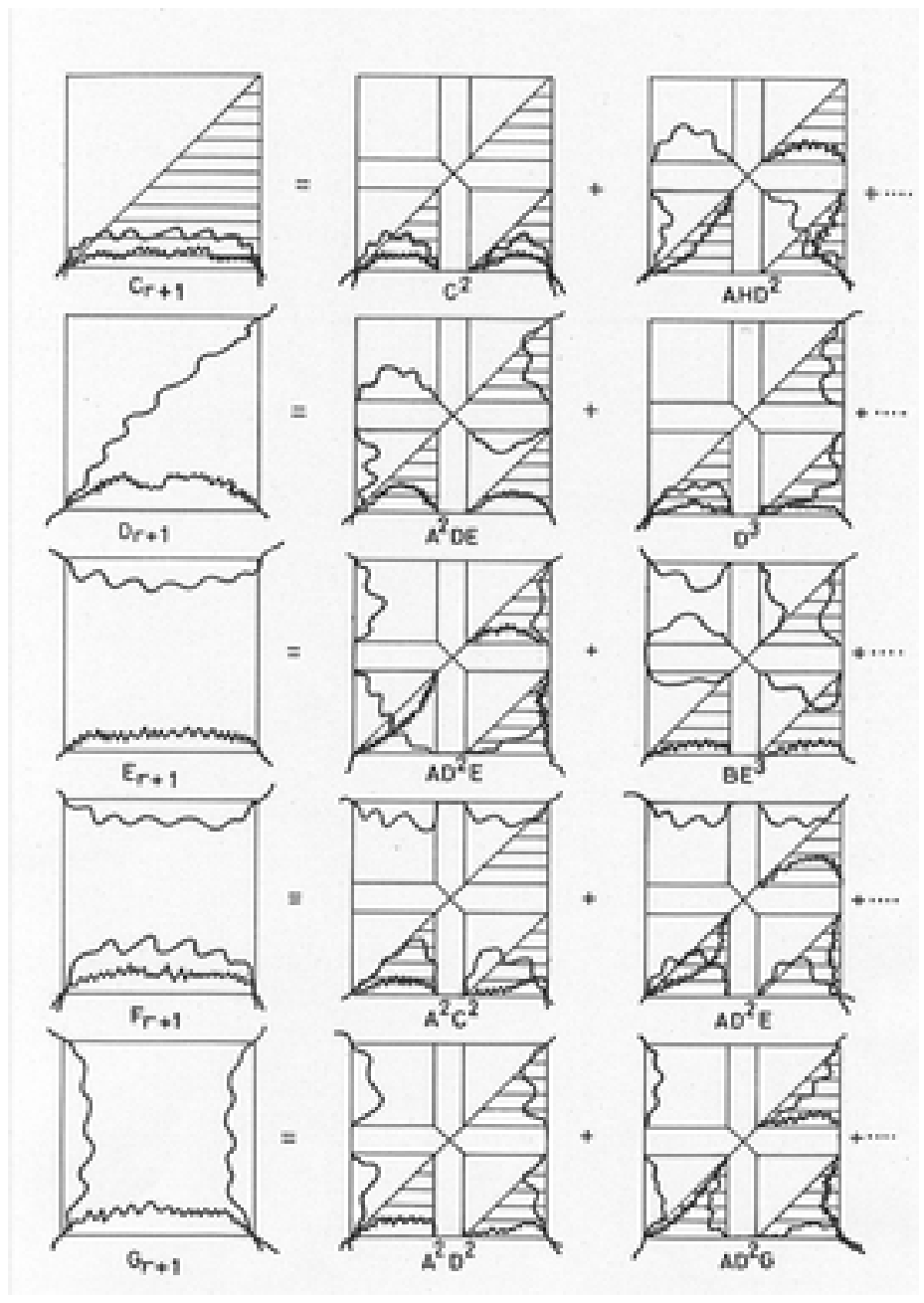,width=12cm}
\caption{ }
\end{center}
\end{figure}
\newpage
\begin{figure}[H]
\begin{center}
\epsfig{file=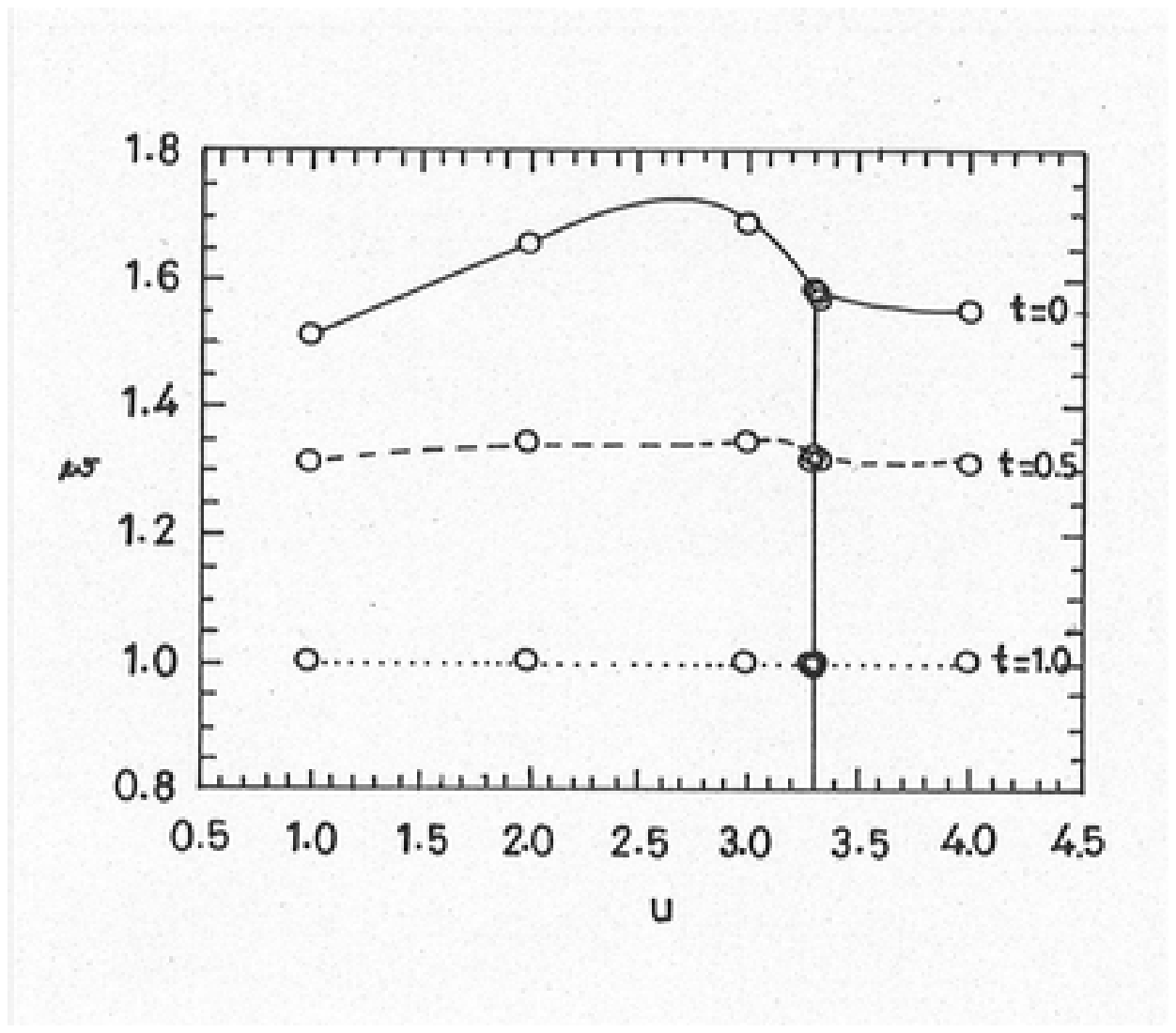,width=12cm}
\caption{ }
\end{center}
\end{figure}
\end {document}